\documentclass[]{spie}  
\usepackage[]{graphicx}
\usepackage{amsmath,amssymb,mathtools,url}
\title{\Large \bf Experimental analysis of the achromatic performance of a\\ vector vortex coronagraph} 
\usepackage{array}
\usepackage{hyperref}
\newcolumntype{P}[1]{>{\centering\arraybackslash}p{#1}}

\usepackage[table]{xcolor}

\author{
Garreth~Ruane\supit{a}, Eugene~Serabyn\supit{a}, Camilo Mejia Prada\supit{a}, Wesley~Baxter\supit{a}, Eduardo~Bendek\supit{a}, Dimitri~Mawet\supit{b,a}, and A~J~Eldorado~Riggs\supit{a} \\
\supit{a}Jet Propulsion Laboratory, California Institute of Technology, 4800 Oak Grove Dr., \\Pasadena, CA 91109, USA\\
\supit{b}Department of Astronomy, California Institute of Technology, 1200 E. California Blvd., \\Pasadena, CA 91125, USA
}

\authorinfo{Send correspondence to gruane@jpl.nasa.gov}

\graphicspath{{Figures/}}
 
\begin{document} 
  \maketitle 

\begin{abstract}
The vector vortex coronagraph is an instrument designed for direct detection and spectroscopy of exoplanets over a broad spectral range. Our team is working towards demonstrating contrast performance commensurate with imaging temperate, terrestrial planets orbiting solar-type stars using the High Contrast Imaging Testbed facility at NASA’s Jet Propulsion Laboratory. To date, the best broadband performance achieved is $\sim$10$^{-8}$ raw contrast over a bandwidth of $\Delta\lambda/\lambda$=10\% in the visible regime (central wavelengths of 550~nm-750~nm), while monochromatic tests yield much deeper contrast ($\sim$10$^{-9}$ or better). In this study, we analyze the main performance limitations on the testbeds so far, focusing on the quality of the focal plane mask manufacturing. We measure the polarization properties of the masks and the residual electric field in the dark hole as a function of wavelength. Our results suggest that the current performance is limited by localized defects in the in the focal plane masks. A new generation of masks is under test that have fewer defects and promise performance improvements. 
\end{abstract}



\section{INTRODUCTION}
\label{sec:intro} 

Coronagraph testbeds have demonstrated contrast levels commensurate with imaging Earth-like planets in the visible regime\cite{Patterson2019,Seo2019}, paving the way for future mission concepts such as the Habitable Exoplanet Observatory (HabEx)\cite{HabEx_finalReport} and the Large UV/Optical/IR Surveyor (LUVOIR)\cite{LUVOIR_finalReport}. While the best contrast performance has been achieved using relatively simple Lyot coronagraphs\cite{Seo2019}, there is significant interest in developing coronagraph masks with the potential to improve performance over a variety of metrics\cite{Ruane2018_metrics}. Vortex coronagraphs\cite{Mawet2005,Foo2005} can theoretically achieve similar contrast performance as Lyot coronagraphs, but over a broader spectral bandwidth with higher throughput at small angular separations and potentially with a larger high-contrast field-of-view, while also relaxing low-order wavefront error requirements\cite{Ruane2018_JATIS}. 

Ground-based instruments operating in the infrared make use of vortex coronagraphs with annular groove phase masks (i.e. subwavelength gratings) in their focal plane\cite{Mawet2005}. Recent observations with the NIRC2 instrument at W.M. Keck Observatory in the $L^\prime$ and $M_s$~bands (3.4-4.8~$\mu$m)\cite{Xuan2018,Guidi2018,Mawet2019_epseri,Uyama2020,Wang2020} and the NEAR instrument on the Very Large Telescope (VLT) in $N$~band (10-12.5~$\mu$m)\cite{Maire2020,Wagner2020} have demonstrated several advantages, including broadband starlight suppression, a small inner working angle, and high throughput, but are limited to raw contrasts that are at least five orders of magnitude worse than the HabEx and LUVOIR requirements. 

Vortex coronagraph masks are a key component in the context of technology development for future space telescopes. The PICTURE-C high-altitude balloon mission\cite{Cook2015,Mendillo2019} has demonstrated a vortex coronagraph in flight and the proposed Coronagraphic Debris Exoplanet Exploring Pioneer (CDEEP) mission\cite{Maier2020} aims to use a vortex coronagraph for visible scattered light imaging of circumstellar disks from a low-Earth orbit. While each of these will provide critical operational and system-level experience, the contrast requirements are also relaxed compared to the HabEx and LUVOIR mission concepts. 

The visible-light vortex coronagraph instruments envisioned for future space telescopes baseline liquid crystal vector vortex waveplates\cite{Mawet2009,Serabyn2019} because they have achieved the best broadband performance to date. For bandwidths of $\Delta\lambda/\lambda$=10\%, a few testbeds have reached raw contrasts of $\sim$10$^{-8}$ at central wavelengths of 550-750~nm, while monochromatic tests yield significantly deeper contrast ($\sim$10$^{-9}$ or better)\cite{Serabyn2013,SerabynTDEM1,SerabynTDEM2,MejiaPrada2019,LlopSayson2020}. The primary motivation for this work is to investigate the limitations in the broadband performance. In the following, we recall the theory of vector vortex coronagraphs, characterize representative vector vortex masks using microscopic imaging polarimetry, and show the contrast performance on one of the four coronagraph testbeds operating in the High Contrast Imaging Testbed facility (HCIT) at NASA’s Jet Propulsion Laboratory (JPL). Based on our laboratory results, we attribute the broadband contrast limitations to localized defects in the vortex focal plane masks that are not easily corrected using conventional wavefront control techniques. 

\section{Vector vortex coronagraph theory}

In this section, we recall the basic principles of vector vortex coronagraphs that are relevant to this study. 

\subsection{An ideal vector vortex mask} 

A vortex coronagraph applies a phase shift at the focal plane mask (FPM) of the form $\exp(il\theta)$, where $l$ is an even integer known as the charge and $\theta$ is the azimuthal angle with the image of the star centered at the origin\cite{Mawet2005,Foo2005,Ruane2018_JATIS}. The phase pattern causes the starlight to diffract outside of a downstream pupil stop, whereas the light from off-axis sources, such as an exoplanet, passes through the system with minor losses. 

A vector vortex mask achieves the desired transmission by manipulating the geometric phase\cite{Mawet2009}. The vector vortex masks described herein are manufactured using photo-aligned liquid crystal polymers to produce a half-wave plate with a spatially-variant fast axis. Such a device may be represented in terms of Jones matrix $\mathbf{M}$, where 
\begin{equation}
\left[ \begin{matrix}
   U^\prime_x \\
   U^\prime_y \\
\end{matrix} \right]=\mathbf{M}\left[ \begin{matrix}
   U_x \\
   U_y \\
\end{matrix} \right]
\end{equation}
and $U_x$ and $U_y$ are the $x$ and $y$ polarized field components. Both the Jones matrix and the field are generally functions of position. An ideal half-wave plate with fast axis orientation angle $\chi$ has the Jones matrix
\begin{equation}
\mathbf{M}=
\mathbf{R}(\chi)
\left[ \begin{matrix}
   1 & 0  \\
   0 & -1  \\
\end{matrix} \right]
\mathbf{R}(-\chi)
=\left[ \begin{matrix}
   \cos 2\chi  & \sin 2\chi   \\
   \sin 2\chi  & -\cos 2\chi   \\
\end{matrix} \right],
\end{equation}
where $\mathbf{R}(\alpha)$ represents a rotation by angle $\alpha$:
\begin{equation}
    \mathbf{R}(\alpha)=\left[ \begin{matrix}
  \cos \alpha  & -\sin \alpha   \\
  \sin \alpha  & \cos \alpha   \\
\end{matrix} \right].
\end{equation}
Converting to the circular polarization basis, the Jones matrix becomes
\begin{equation}
\mathbf{M}_\circlearrowright=\left[ \begin{matrix}
   1 & i  \\
   1 & -i  \\
\end{matrix} \right] \left[ \begin{matrix}
   \cos 2\chi  & \sin 2\chi   \\
   \sin 2\chi  & -\cos 2\chi   \\
\end{matrix} \right] \left[ \begin{matrix}
   1 & i  \\
   1 & -i  \\
\end{matrix} \right]^{-1}
=\left[ \begin{matrix}
   0 & e^{i2\chi}  \\
   e^{-i2\chi} & 0  \\
\end{matrix} \right].
\end{equation}
The result is effectively a phase-only mask applied to each circular polarization state: 
\begin{equation}
\left[ \begin{matrix}
   U^\prime_R \\
   U^\prime_L \\
\end{matrix} \right]=\mathbf{M}_\circlearrowright \left[ \begin{matrix}
   U_R \\
   U_L \\
\end{matrix} \right]
= \left[\begin{matrix}
   0 & {{e}^{i2\chi}}  \\
   {{e}^{-i2\chi}} & 0  \\
\end{matrix} \right] \left[ \begin{matrix}
   U_R\\
   U_L\\
\end{matrix} \right],
\end{equation}
where $U_R$ and $U_L$ are the right- and left-handed circular polarization field components, respectively. And, the applied phase function is $\Phi=\pm 2\chi $, where the sign depends on the handedness of the incident polarization and the phase shift depends on the local orientation angle of the fast axis. A vector vortex mask has $\chi=l\theta/2$, where $l$ is the charge. 

\subsection{Leakage due to imperfect retardance} 

In reality, it's not possible to manufacture a perfectly achromatic half-wave plate. For example, if the vector phase mask has an uniform retardance error, the Jones matrix in the linear polarization basis becomes
\begin{equation}
\mathbf{M}=\mathbf{R}(\chi)
\left[ \begin{matrix}
   1 & 0  \\
   0 & e^{i(\pi+\epsilon_V)}  \\
\end{matrix} \right]
\mathbf{R}(-\chi),
\end{equation}
where $\epsilon_V$ is the retardance error in the mask. Converting to the circular polarization basis as above: 
\begin{equation}
\mathbf{M}_\circlearrowright
=c_V\left[ \begin{matrix}
   0 & e^{i2\chi}  \\
   e^{-i2\chi} & 0  \\
\end{matrix} \right] + 
c_L\left[ \begin{matrix}
   1 & 0  \\
   0 & 1  \\
\end{matrix} \right],
\end{equation}
where $c_V$ and $c_L$ are constants. The second term results in a stellar leakage whose phase is unchanged by the mask and whose intensity is $|c_L|^2$ times that of the incident beam, where $|c_L|^2=\sin^2(\epsilon_V/2)$. The fraction of the total power that transfers into the optical vortex beam is $|c_V|^2=\cos^2(\epsilon_V/2)$. For small retardance errors (i.e. $\epsilon_V\ll$1~rad), $|c_L|^2\approx\epsilon_V^2/4$ and $|c_V|^2\approx1-\epsilon_V^2/4$. Generally, $\epsilon_V$ is a function of wavelength and can limit the usable bandwidth of the coronagraph. 

\begin{figure}[t]
    \centering
    \includegraphics[width=\linewidth]{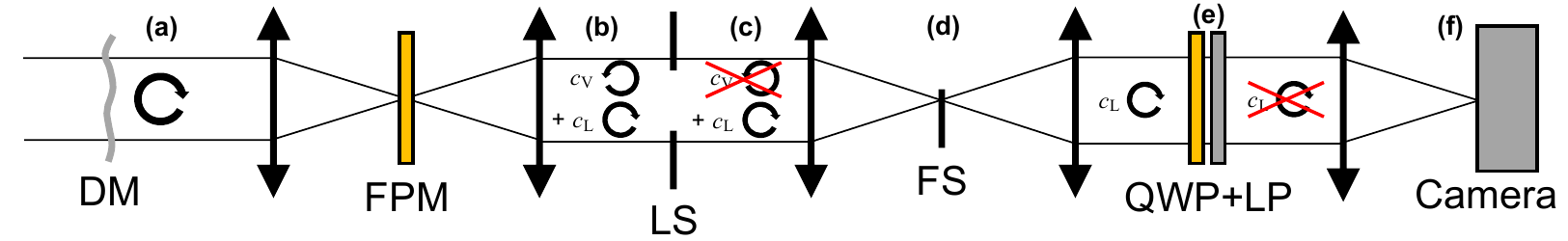}
    \caption{Schematic of a vector vortex coronagraph with polarization filtering. \textbf{(a)}~The light from the star and planet are circularly polarized before the coronagraph entrance pupil, which is typically near the deformable mirror~(DM). \textbf{(b)}~After passing through the vector vortex focal plane mask (FPM), the light has two polarization components: the vortex term flips handedness (with fractional energy $|c_V|^2$) and the leakage term which doesn't change polarization (fractional energy is $|c_L|^2$). The vortex term for the starlight takes on the desired phase term of $\exp(il\theta)$, while the stellar leakage term doesn't have the vortex phase. On the other hand, the planet light is not converted into an optical vortex in either polarization component because it is not aligned to the phase singularity at the center of the FPM. \textbf{(c)}~The stellar vortex term is mostly blocked by the Lyot stop (LS) with a residual that depends on the wavefront error. Both the stellar leakage term and the planet light pass through the LS without being suppressed by the coronagraph masks. \textbf{(d)}~The field stop (FS) is used to block bright regions of the image plane that are outside of the high-contrast field of view, but also doesn't reduce the impact of the leakage term. \textbf{(e)}~A circular analyzer made up of a quarter wave plate (QWP) and linear polarizer (LP) cancels the leaked polarization components (i.e. the original polarization state) for the planet and star. However, the extinction of the analyzer is imperfect and wavelength dependent. \textbf{(f)}~The residual starlight at the camera depends on the wavefront error, the magnitude of the leakage term (i.e. the bulk retardance error), and the extinction of the circular analyzer. After minimizing these three terms, we find that the next most dominant contrast limitation is localized defects in the FPM.}
    \label{fig:schematic}
\end{figure}

\subsection{Polarization filtering} 

A method to improve the starlight reject with an imperfect vector vortex mask is to block the leaked term by circularly polarizing the beam before it reaches the mask and then using a circular analyzer that only allows the orthogonal polarization to reach the final detector (see Fig.~\ref{fig:schematic})\cite{Serabyn2013}. This blocks the light that does not have the intended vortex phase pattern. To illustrate this, we represent a perfect linear polarizer (LP) and quarter waveplate (QWP) as:
\begin{equation}
\mathbf{J_\text{P}}(\theta)=
\mathbf{R}(\theta)
\left[ \begin{matrix}
   1 & 0  \\
   0 & 0  \\
\end{matrix} \right]
\mathbf{R}(-\theta),
\end{equation}
\begin{equation}
\mathbf{J_\text{Q}}(\theta)=
\mathbf{R}(\theta)
\left[ \begin{matrix}
   1 & 0  \\
   0 & i  \\
\end{matrix} \right]
\mathbf{R}(-\theta),
\end{equation}
where $\mathbf{J_\text{P}}$ and $\mathbf{J_\text{Q}}$ are the Jones matrices for the LP and QWP, respectively, and $\theta$ is the axis of transmission for the LP and fast axis angle for the QWP. Then, the resulting Jones matrix of the full system in the linear polarization basis is
\begin{equation}
    \mathbf{J_\text{sys}} = \mathbf{J_\text{P}}\left(\frac{\pi}{2}\right) \mathbf{J_\text{Q}}\left(\frac{\pi}{4}\right)
    \mathbf{M} \;
    \mathbf{J_\text{Q}}\left(\frac{-\pi}{4}\right) \mathbf{J_\text{P}}(0) 
    =c_V \left[ \begin{matrix}
    0  & 0   \\
    e^{i2\chi}  & 0   \\
    \end{matrix} \right].
\end{equation}
The leakage term is blocked while the output beam has the intended phase and is linearly polarized.

In past analytical work\cite{Ruane2019_scalarVC}, we also included terms that represent imperfections in the LP extinction and QWP retardance, which may also limit performance. However, we will show that the bulk properties of the LP and QWP are not dominant our current experimental results. Rather, in the following, we find that local manufacturing errors in the vortex mask are likely limiting the contrast performance of the coronagraph after polarization filtering. 

\section{Vortex mask characterization}

Over the past 10 years, our team at JPL has been working with manufacturers to improve the quality of vector vortex masks, and through this process has acquired over 100 liquid crystal masks for the visible regime. The quality of the masks has been consistently improving with each generation. To characterize the masks, we use a Mueller Matrix Imaging Polarimeter (MMIP; Axometrics Axostep), which provides the 4$\times$4 Mueller matrix for each pixel in a microscope image of the mask as a function of wavelength over 400-800~nm.  For the sake of this study, we present results for two masks manufactured by ``BEAM Co." circa 2018. 

\begin{figure}[t]
    \centering
    \includegraphics[width=\linewidth]{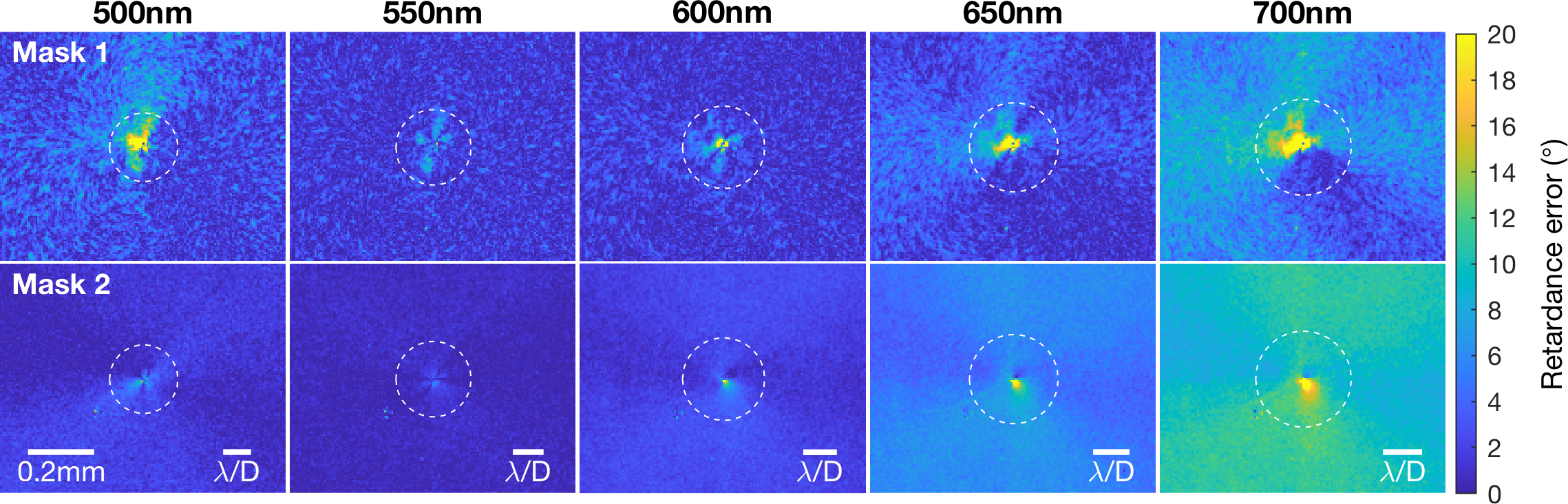}
    \caption{Retardance error (absolute difference from 180$^\circ$) near the center of the two vector vortex masks (`Mask 1' and `Mask 2'). The regions shown are 691$\times$864~$\mu$m with 5.4~$\mu$m per pixel and centered on the phase singularity. The $\lambda/D$ scale bar is specific to our vortex coronagraph testbed, which had a focal ratio of 170 at the FPM. The dashed white circle indicates the location of the first zero in the Airy diffraction pattern for scale (1.22$\lambda/D$ in radius, which corresponds to 114~$\mu$m at the central wavelength of 550~nm). }
    \label{fig:retardance_grid}
\end{figure}

Figure~\ref{fig:retardance_grid} shows the retardance errors measured by the MMIP at representative wavelengths for the two masks (denoted `Mask 1' and `Mask 2'). Each mask was designed for $l=4$, a central wavelength of $\lambda_0$~=~550~nm, and a goal bandwidth of $\Delta\lambda/\lambda_0 \approx 20\%$. However, we studied the masks over a larger spectral range to empirically determine the wavelengths where the mask performs best and understand the degradation away from the central wavelength. The most prominent feature in the retardance error maps was typically near the center phase singularity. For reference, the dashed white circle in Fig.~\ref{fig:retardance_grid} is centered on the singularity and represents the size of the Airy disk (i.e. 1.22$\lambda/D$ radius) at the FPM in our coronagraph testbed, which has a focal ratio of 170. The testbed will be presented in more detail in the next section. Each mask had a small, metallic, circular spot that covered the center defect (15~$\mu$m and 10~$\mu$m in diameter for Masks 1 and 2, respectively.) In theory, the impact of the opaque metal spot and retardance errors within the central $\sim\lambda/D$ is relatively small\cite{Ruane2018_JATIS}. The features that have a dominant impact on the raw contrast are the retardance errors outside of, but near, the central $\sim\lambda/D$ and significant local defects in the mask within $\sim$20-30~$\lambda/D$ of the center. 

\begin{figure}[t]
    \centering
    \includegraphics[width=0.45\linewidth]{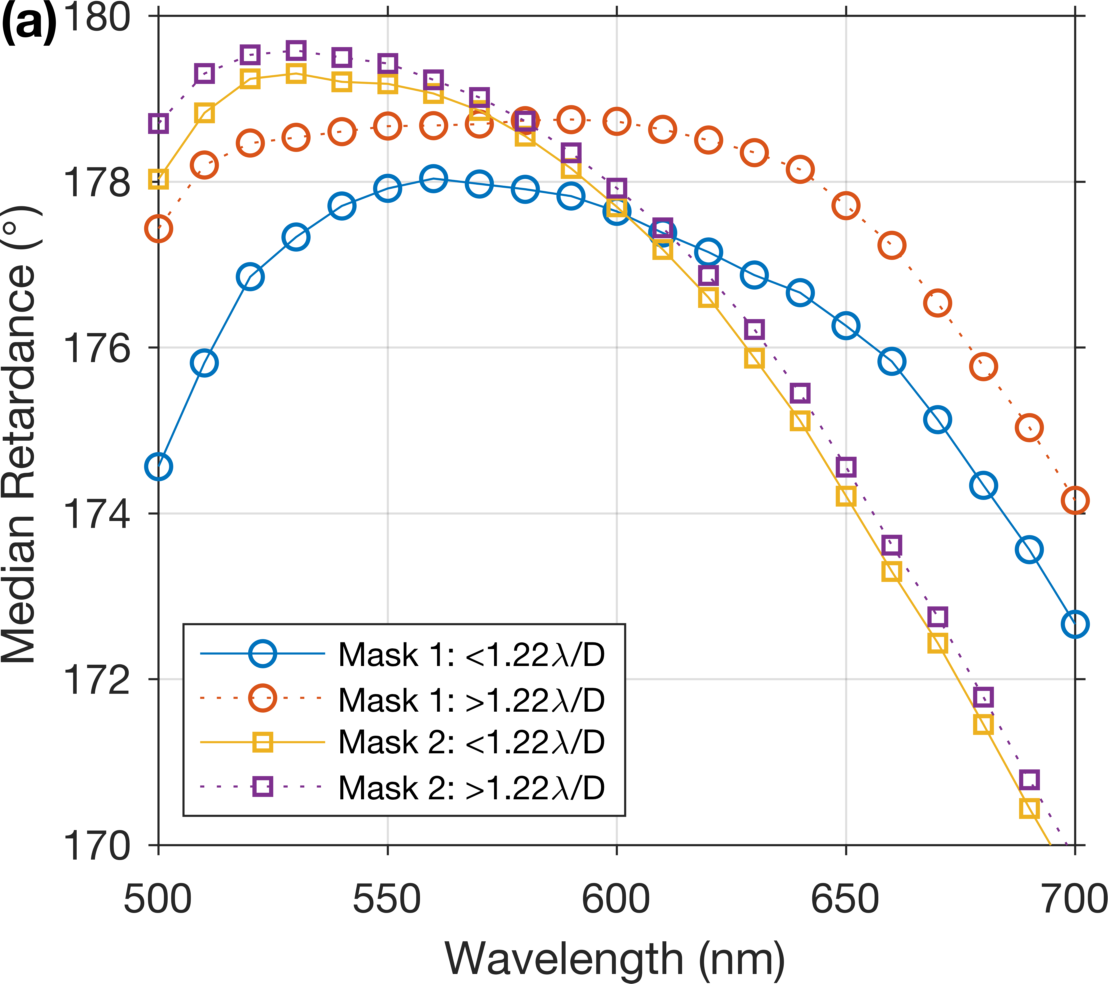}\hspace{5mm}
    \includegraphics[width=0.45\linewidth]{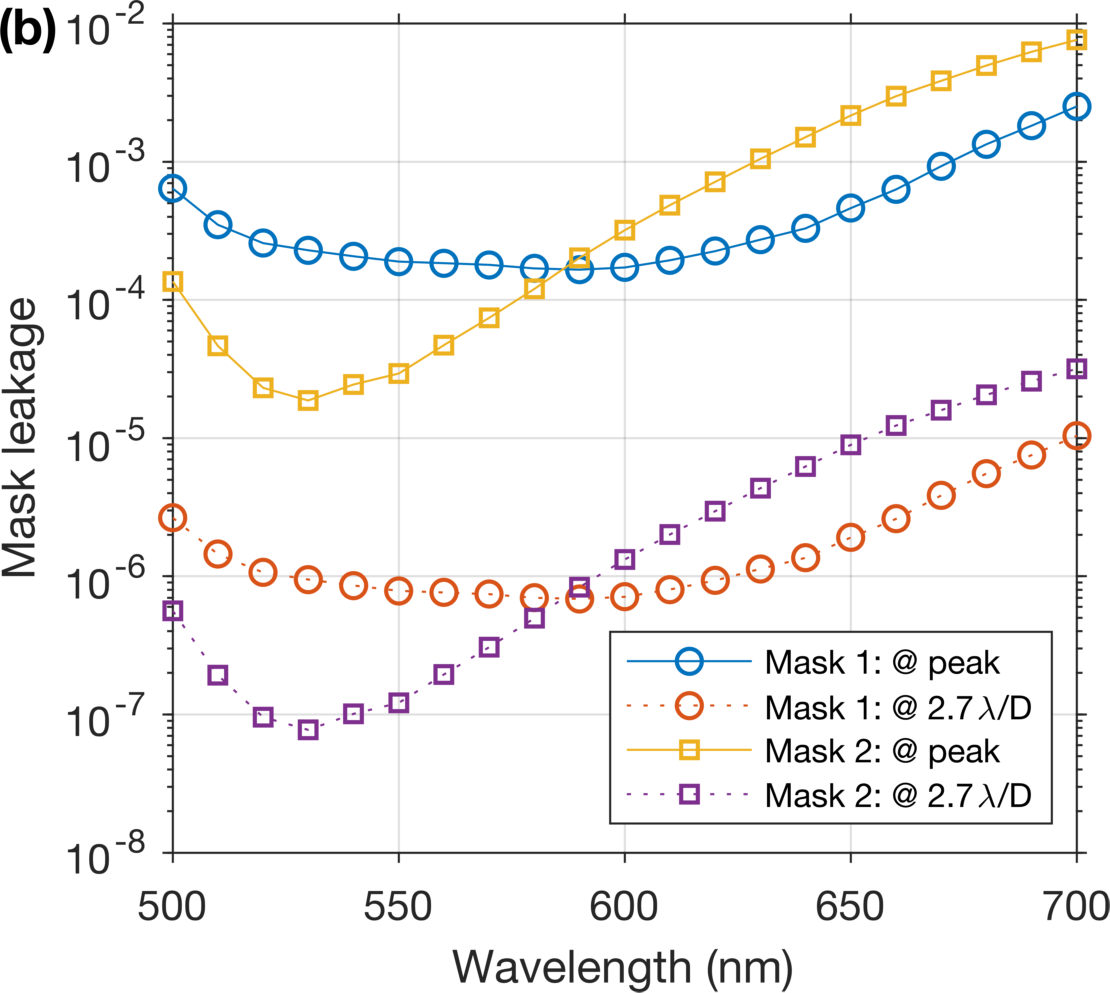}
    \caption{\textbf{(a)}~Median retardance as a function of wavelength for the two masks for separations inside and outside of 1.22~$\lambda/D$ (i.e. the white circle in Fig.~\ref{fig:maskleakage}). \textbf{(b)}~The leaked intensity based on the analytical calculation of $|c_L|^2$ using the bulk retardance error outside of the Airy disk. The leak results in an Airy pattern in the final image plane with diffraction rings that contaminate the dark hole. The second Airy ring located at 2.7~$\lambda/D$ has a relative intensity of 4$\times 10^{-3}$ with respect to the peak. This calculation is solely referring to the leakage from the mask itself and does not include the leakage reduction from polarization filtering with a circular analyzer. }
    \label{fig:maskleakage}
\end{figure}

We computed the median error both inside and outside of the Airy disk (see Fig.~\ref{fig:maskleakage}a). Mask~1 has larger errors within the central region, but the bulk retardance error is relatively flat from roughly 500~nm to 650~nm. Mask~2 has smaller errors near the center and is closer to 180$^\circ$ retardance within 500-570~nm, but has larger errors at longer wavelengths than Mask~1. To estimate the leakage term due to the bulk retardance error, we used the outer retardance value and the analytical model for the leaked energy derived above: $|c_L|^2=\sin^2(\epsilon_V/2)$, where $\epsilon_V$ is the retardance error (see Fig.~\ref{fig:maskleakage}b). The leaked term appears in the final image plane as an Airy pattern with a peak intensity that is $|c_L|^2$ times fainter than the case without the vortex mask in the system. Since we wish to image exoplanets that are resolved from their host star, the region of interest is outside of the central lobe of the Airy disk. For example, the peak of the second Airy ring is 2.7~$\lambda/D$ away from the star position and has a relative intensity of 4$\times 10^{-3}$ with respect to the peak. Thus, the second airy ring of the leaked light is expected to appear at a contrast of $\sim10^{-6}$ from 500~nm to 650~nm for Mask~1, and as small as $\sim10^{-7}$ at the optimal wavelength for Mask~2. This leakage term is further reduced by polarization filtering with the circular analyzer, which typically has an extinction on the order of 10$^{-4}$-10$^{-3}$ and thus brings the leakage terms down to a contrast level of 10$^{-10}$-10$^{-9}$ within the design bandwidth. 

In theory, the bulk properties of the masks meet, or at least are close to meeting, the requirements for a polarization-filtered vector vortex coronagraph for future space telescopes. However, we also measure small scale variations in the retardance, which are especially prominent in Mask~1, as well as small defects in both the retardance and transmittance. In the following, we present results from our vortex coronagraph testbed that suggest that these features are limiting the contrast in practice. 

\section{Coronagraph Testbed Results}

In this section, we introduce the HCIT vortex coronagraph testbed, the wavefront control methods we apply, and the resulting contrast performance in narrowband and broadband light. 

\subsection{The vortex coronagraph testbed}

The vortex coronagraph testbed was primarily used to test liquid crystal vector vortex coronagraph masks. This testbed was installed in one of HCIT's vacuum chambers, but has since been removed to make way for a new generation of testbeds known as the Decadal Survey Testbeds\cite{DSTroadmap}. 

\begin{figure}[t]
    \centering
    \includegraphics[width=\linewidth]{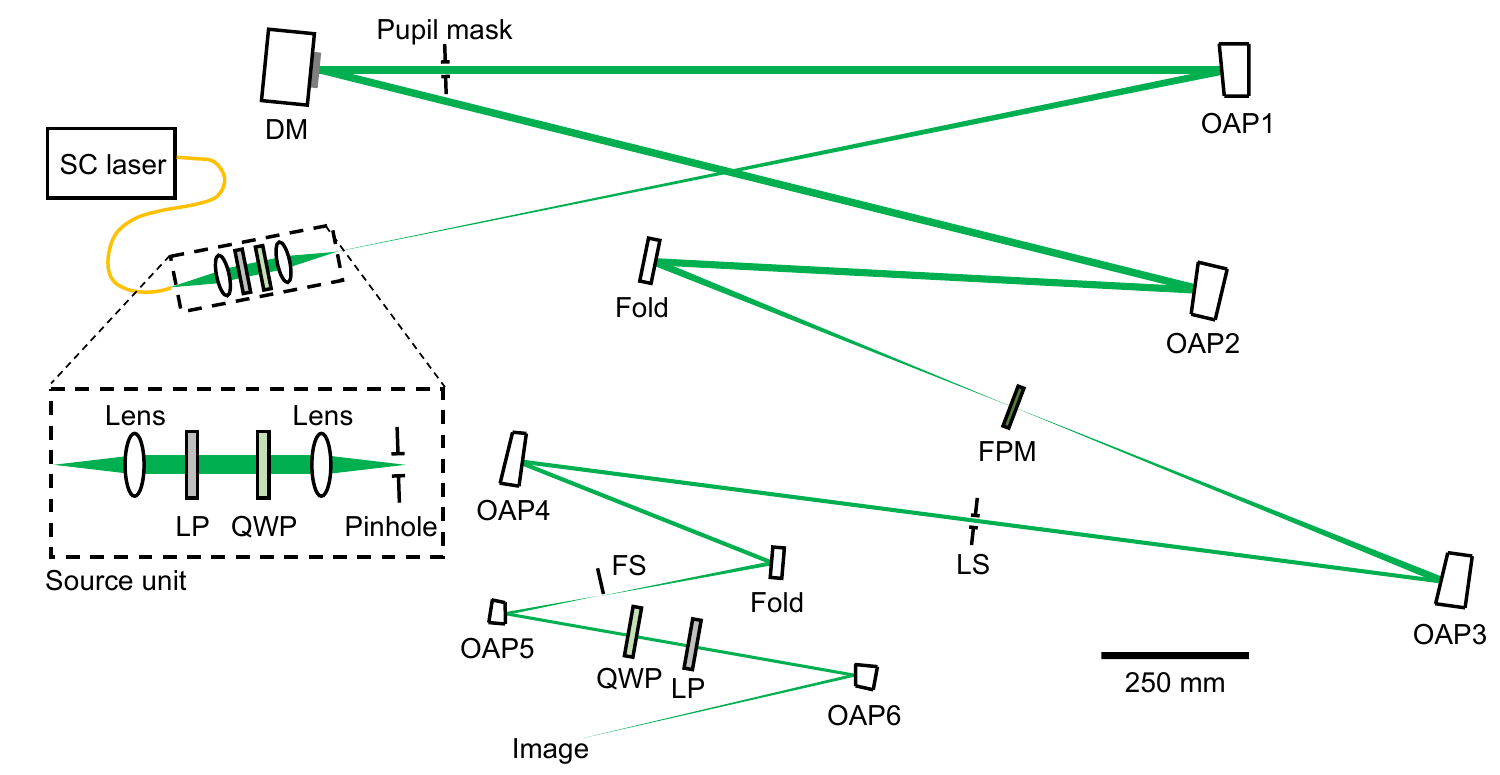}
    \caption{Schematic of the vortex coronagraph testbed. SC: Supercontinuum. LP: Linear polarizer. QWP: Quarter wave plate. OAP: Off axis parabola. DM: Deformable mirror. FPM: Focal plane mask. LS: Lyot stop. FS: Field stop. }
    \label{fig:gpct_layout}
\end{figure}

Figure~\ref{fig:gpct_layout} shows the vortex coronagraph testbed layout. In the ``source unit," light from a fiber-coupled supercontinuum (SC; NKT SuperK) laser was collimated by a lens, circularly polarized using a LP and QWP, and focused onto a pinhole (3~$\mu$m in diameter) to define a quasi-point source. After the pinhole, the beam was re-collimated by the first off-axis parabolic mirror (OAP1; $f$=1.5m) and passed through a circular pupil mask (9~mm diameter). The beam then reflected off the deformable mirror (DM), was focused onto the focal plane mask (FPM) by OAP2 ($f$=1.5m), and was re-collimated by OAP3 ($f$=0.76m). The Lyot stop (LS; 3.9~mm diameter) was 86\% of the re-imaged pupil (4.5~mm diameter). The vortex FPM diffracted most of the starlight outside of the LS and OAP4 ($f$=0.76m) focused the residual light at the field stop (FS). The FS was a simple knife edge that blocked more than half of the image plane including the core of the leakage term. After the FS, OAP5 ($f$=0.15m) collimated the beam which then passed through the QWP and LP that made up the circular analyzer. OAP6 ($f$=0.46m) focused the beam onto the final imaging camera (Andor Neo sCMOS). The camera had a pixel pitch of 6.5~$\mu$m. All of the components were held at the vacuum chamber pressure of 0.1~mTorr except the SC laser source, which was outside of the vacuum chamber and used a fiber feedthrough, and the camera, which was inside of its own vented chamber with a transmissive window. 

The DM was a Boston Micromachines Kilo-DM with 952 actuators with inter-actuator spacing of 300~$\mu$m located 63~mm downstream of the pupil mask. The circular beam illuminated an area that was 29.8 actuators across. The DM was controlled using a low-noise 16-bit electronics manufactured by Teilch\cite{Bendek2020} and set up to provide a range of 100~V and a surface height resolution estimated to be below the achieved contrast floor\cite{Ruane2020_LSB}. The electronics were cooled using a Peltier device that transferred heat into water cooling lines held at 17$^\circ$C. 

We tested a number of vortex masks and obtained slightly different results with each. The contrast performance and primary contrast limitations were similar between the masks used in this work and previous experiments. In the following, we present results from our most studied mask (`Mask 1'), which is not the best quality mask by any measure. The main motivation for this work was to determine what is limiting the broadband performance of the masks in general, using `Mask 1' as an example.  

\subsection{Wavefront control}

Prior to installing the DM on the testbed, it was calibrated using a Fizeau interferometer (Zygo Verifire). For the interferometric calibration, the DM and interferometer were kept at atmospheric pressure, but the DM was housed inside of a plastic enclosure that was continuously purged with dry air to maintain relative humidity $<$30\% and had a opening for the beam to pass through. The calibration steps were to flatten the DM and measure the displacement per unit voltage (units of nm/V) for each actuator. The voltages needed to flatten the unpowered surface, which is mostly a defocus shape, ranged from 0~V near the edges to 80~V near the center. The final surface error was $<$10~nm~RMS. A typical actuator is displaced by approximately 4~nm/V in the DM's flat state. More information regarding this DM's calibration can be found in Bendek et al. (2020)\cite{Bendek2020}.

After the DM was calibrated and installed on the testbed, we used a phase-retrieval algorithm\cite{Fienup1982} to estimate the phase in the pupil from a sequence of images with the camera in different positions along the beam axis. By retrieving the phase with known patterns applied to the DM, we determined the actuator locations with respect to the beam. This information as well as the phase-retrieval with the DM in its flat state were the key parts of the testbed model used for wavefront control. 

After aligning the coronagraph FPM and LS, we measured the complex field in the final image plane using pair-wise probing (PWP)\cite{Giveon2011,Groff2015} where we applied two pairs of sinc-sine probe patterns to the DM. Then, we used the electric field conjugation (EFC) method to determine the voltage changes needed to minimize the intensity in a chosen region of the final image plane\cite{Giveon2007}. Our iterative implementation of the PWP and EFC algorithms is publicly available in the open-source FALCO package\cite{Riggs2018}.

\begin{figure}[t]
    \centering
    \includegraphics[width=\linewidth]{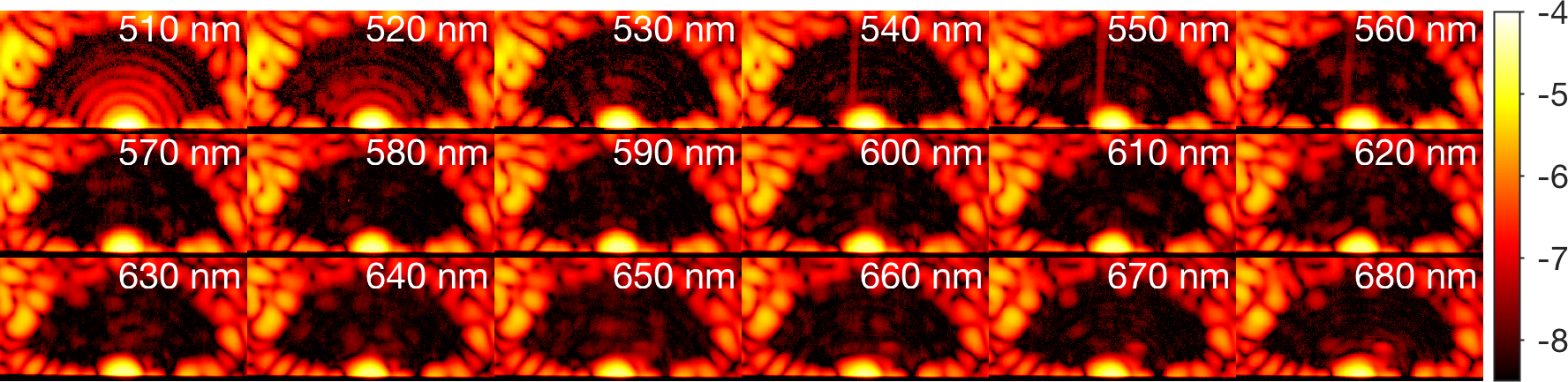}
    \caption{Log normalized intensity in a $\Delta\lambda/\lambda_0$~=~3\% bandwidth versus central wavelength, $\lambda_0$, using `Mask 1.' The dark hole in a 3-9~$\lambda_0/D$ partial annulus with a 140$^\circ$ opening angle. In each panel, the psuedo-star is located at the center of the bottom and the FS edge appears as the horizontal dark strip along the bottom. Polarization leakage is apparent for $\lambda_0<$530~nm (i.e. Airy rings appear in the dark hole). The contrast performance is weakly dependent on wavelength over 530-680~nm.}
    \label{fig:multiwvlDHs_3pctBW}
\end{figure}

\subsection{Narrowband results}

The normalized intensity in the final image is defined as the detector counts divided by the peak counts with the focal plane mask offset such that the beam passes through a part of the mask that is far from the phase singularity. Figure~\ref{fig:multiwvlDHs_3pctBW} shows the resulting normalized intensity after 10 iterations of the PWP+EFC control loop to create a dark hole in a 3-9~$\lambda_0/D$ partial annulus with a 140$^\circ$ opening angle with `Mask~1' and a bandwidth of $\Delta\lambda/\lambda$~=~3\%. Using a variable spectral filter (NKT Varia) at the SC laser source, we repeated this test for $\lambda_0$ ranging from 510~nm to 680~nm in steps of 10~nm. We obtained dark holes with mean normalized intensity $<$10$^{-8}$ for $\lambda_0$~=~530-680~nm. At shorter wavelengths (see 510~nm and 520~nm cases in Fig.~\ref{fig:multiwvlDHs_3pctBW}), an Airy pattern appeared in the dark hole which is indicative of polarization leakage due to bulk retardance errors in the vortex masks and imperfect extinction of the circular analyzer. Recalling Fig.~\ref{fig:maskleakage}b, the mask leakage alone is stronger in the ranges of 500-520~nm as well as $>$650~nm. However, after polarization filtering, the extinction due to the circular analyzer was only $\sim$0.1 in the 500-520~nm range, but significantly better at longer wavelengths where we obtained $<$10$^{-8}$ contrast even with a few degrees of bulk retardance error.

\begin{figure}[t]
    \centering
    \includegraphics[trim={0 0 2.8cm 0},clip,height=0.42\linewidth]{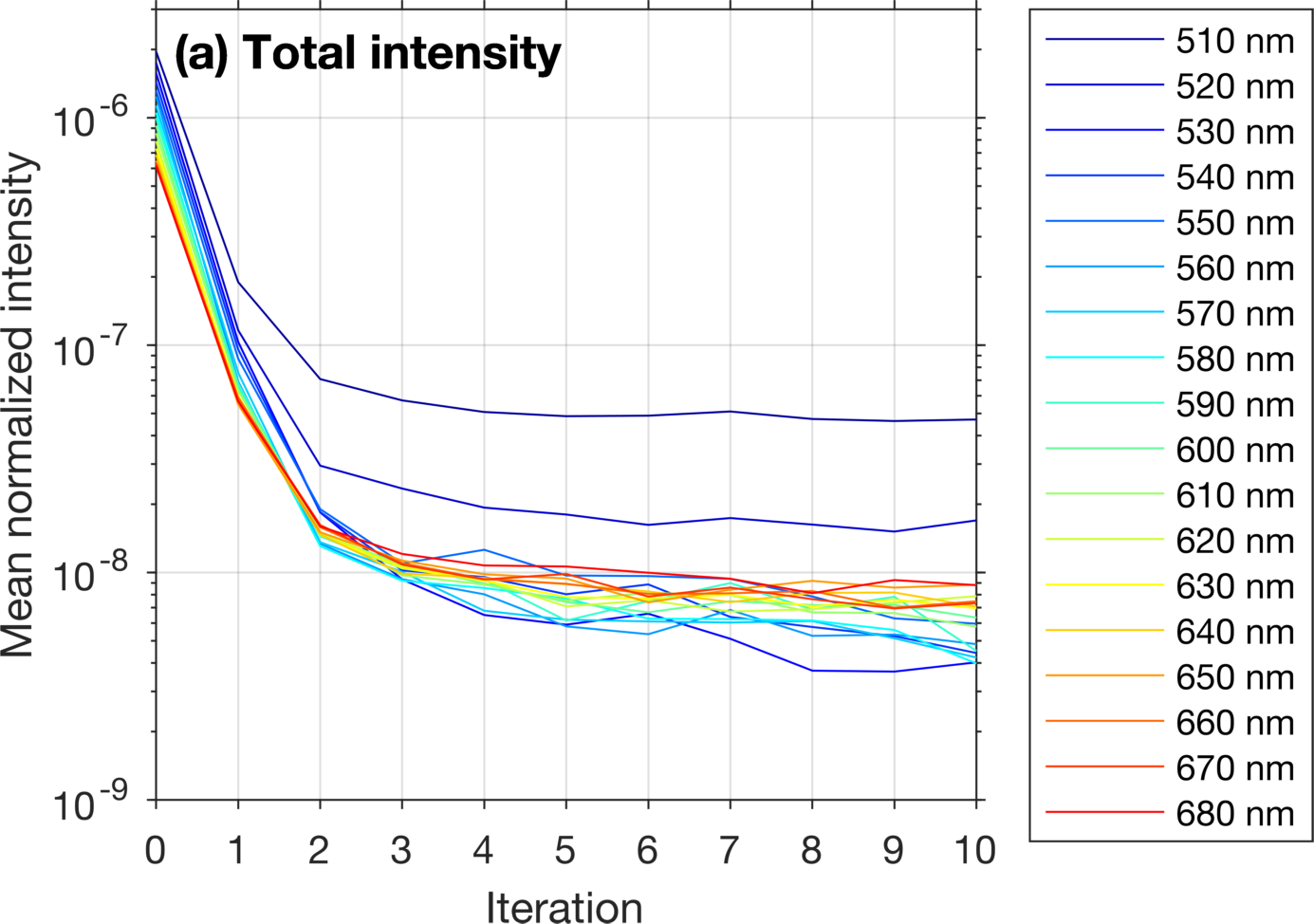}
    \includegraphics[trim={1.2cm 0 0 0},clip,height=0.42\linewidth]{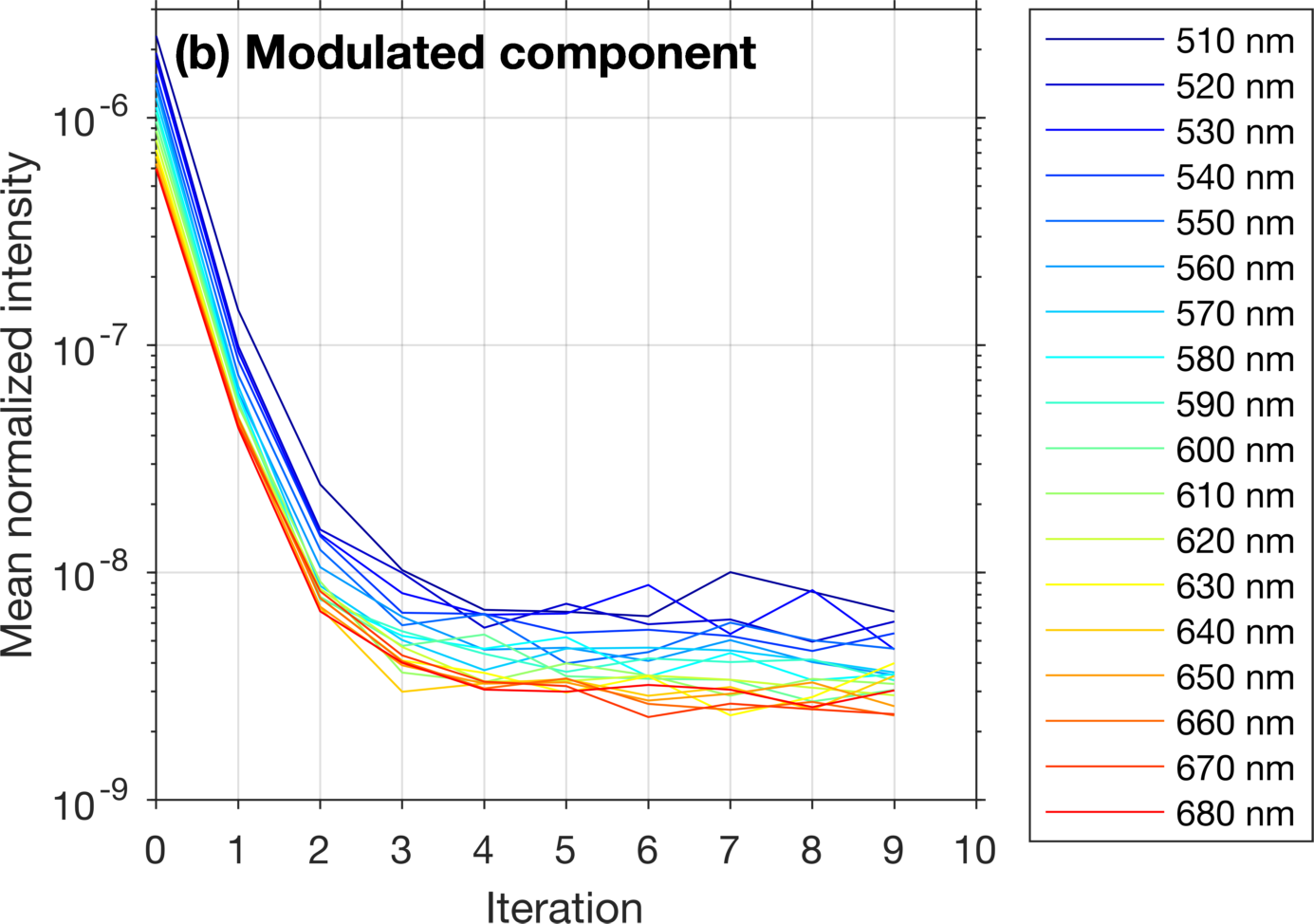}
    \caption{The mean of the \textbf{(a)}~total and \textbf{(b)}~modulated components of the normalized intensity in the dark holes in Fig.~\ref{fig:multiwvlDHs_3pctBW}.}
    \label{fig:multiwvl_normI_vs_it}
\end{figure}

At all wavelengths, the initial normalized intensity was $\sim$10$^{-6}$ with the DM in its flat state. Figure~\ref{fig:multiwvl_normI_vs_it}a shows the normalized intensity versus PWP+EFC iteration for each $\lambda_0$. PWP also provided an estimate of the normalized intensity that modulates with the DM probes\cite{Giveon2011,Groff2015}. Figure~\ref{fig:multiwvl_normI_vs_it}b shows that the modulated component converged to $<$10$^{-8}$ and the floor improved with wavelength as expected for a similar residual surface error on the DM. The difference between the modulated and total intensity is the so-called unmodulated intensity, which captures incoherent light as well as the polarization leakage due to its relatively weak response to the DM probes. Compared to the modulated component, the total intensity followed a trend with wavelength that is consistent with the mask leakage; i.e. it increased for the shortest and longest wavelengths. This indicates that while polarization leakage dominated for the shortest wavelengths, it may not have been negligible for the longest wavelengths. Furthermore, bright spots were visible near the outer portions of the dark hole for the longest wavelengths that did not appear at shorter wavelengths. We attribute these features to defects in the mask that are exposed as the dark hole expanded with $\lambda_0$.

\subsection{Broadband results}

\begin{figure}[t]
    \centering
    \includegraphics[height=0.4\linewidth]{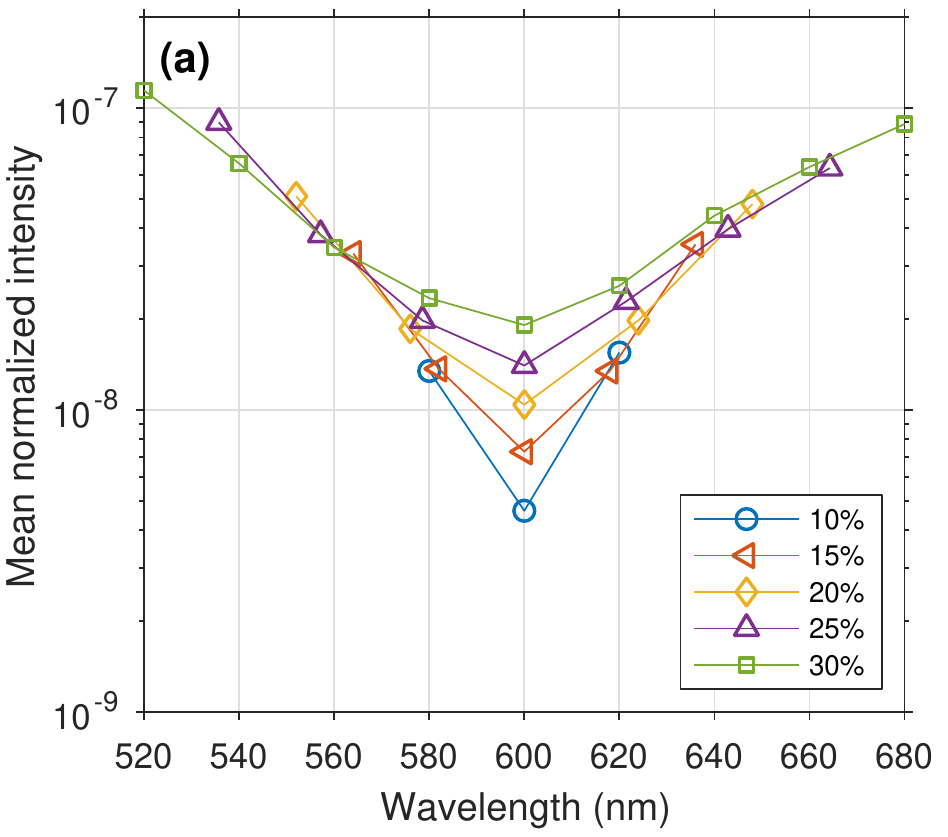}
    \includegraphics[height=0.4\linewidth]{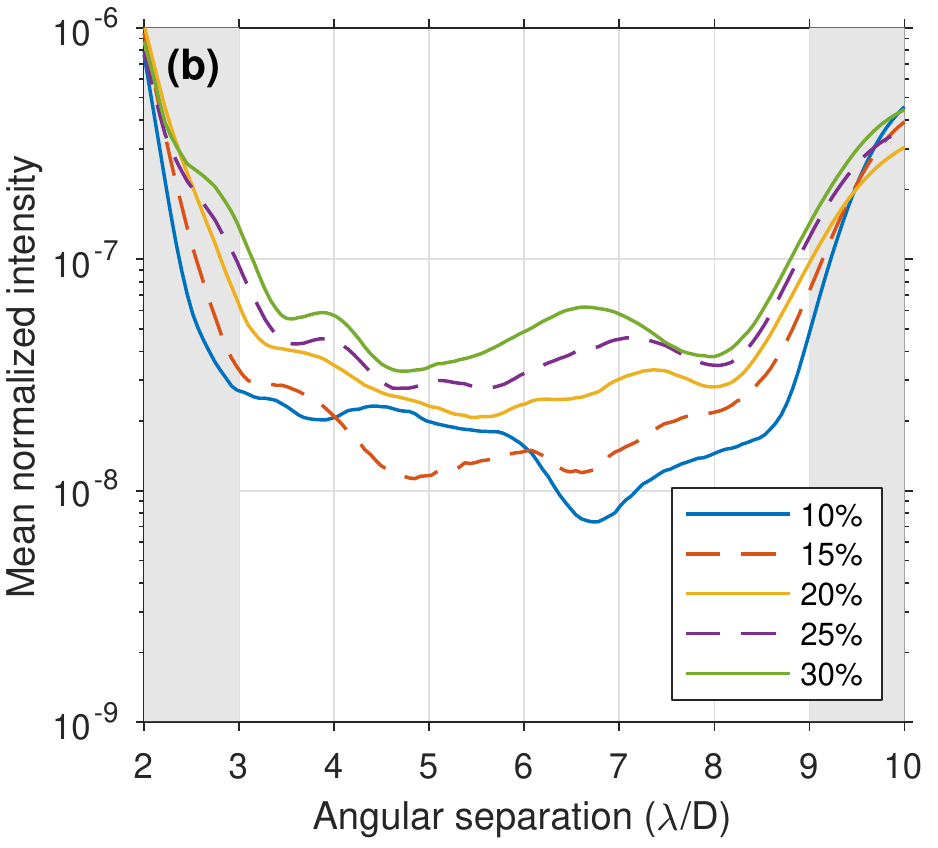}
    \caption{Normalized intensity for simultaneous bandwidths $\Delta\lambda/\lambda$~=~10\% to 30\% in steps of 5\%. \textbf{(a)}~The spatial mean over the dark hole at each sub-band. \textbf{(b)}~The azimuthal and spectral mean versus angular separation from the source.}
    \label{fig:normI_vs_wvl}
\end{figure}

While the narrowband experiments were useful for diagnosing contrast limitations, our ultimate goal was to demonstrate broadband contrast by using PWP+EFC to suppress the pseudo-starlight in multiple sub-bands simultaneously. For broadband wavefront control, we broke the desired bandwidth into a number of sub-bands, performed the PWP for each one, and applied a multi-wavelength EFC method to determine the DM settings that create the broadband dark hole for the same region as Fig.~\ref{fig:multiwvlDHs_3pctBW}. Figure~\ref{fig:normI_vs_wvl} shows the results for bandwidths from $\Delta\lambda/\lambda$~=~10\% to 30\% in steps of 5\%. The markers in Fig.~\ref{fig:normI_vs_wvl}a indicate the central wavelength of the sub-bands. We used 3 sub-bands for the 10\% bandwidth case and up to 9 sub-bands for the 30\% bandwidth case. Overall, the normalized intensity degraded with bandwidth and became flatter as a function of wavelength. Fig.~\ref{fig:normI_vs_wvl}b shows the azimuthal average of the broadband dark hole. The normalized intensity didn't significantly decrease with angular separation. The spatial mean of the broadband normalized intensity was 1.7$\times$10$^{-8}$ for the 10\% bandwidth, 3.3$\times$10$^{-8}$ for the 20\% bandwidth, and 5.4$\times$10$^{-8}$ for the 30\% bandwidth.

\begin{figure}[t]
    \centering
    \includegraphics[width=0.94\linewidth]{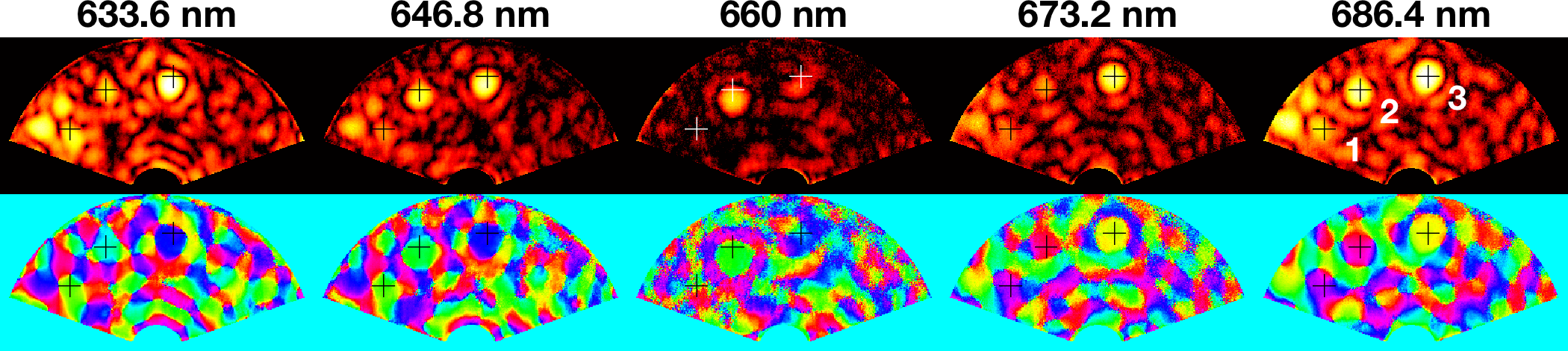}\\
    \includegraphics[width=0.47\linewidth]{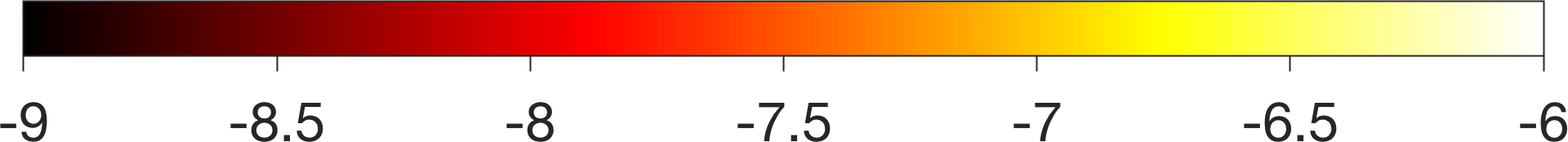}
    \hspace{0.1cm}\includegraphics[width=0.47\linewidth]{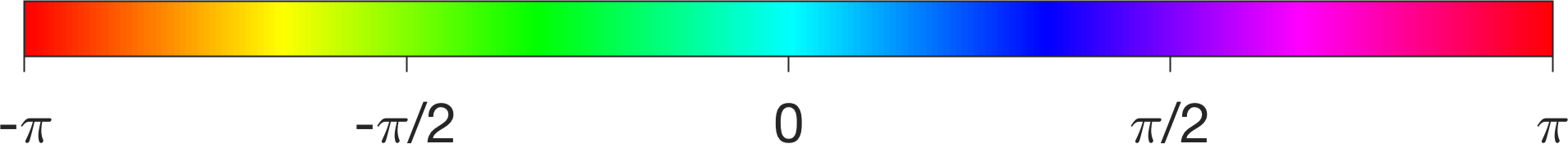}
    \caption{Electric field in the image plane measured with PWP at 5 sub-bands within a 10\% bandwidth centered at $\lambda_0$=~660~nm. The \textbf{(top row)} log normalized intensity and \textbf{(bottom row)} phase correspond to the left and right colorbars, respectively. The values at the crosses in the image as plotted in Fig.~\ref{fig:speckles} for the speckles labeled `1,' `2,' and `3.'}
    \label{fig:Efield_vs_wvl}
\end{figure}

\begin{figure}[t]
    \centering
    \includegraphics[width=0.45\linewidth]{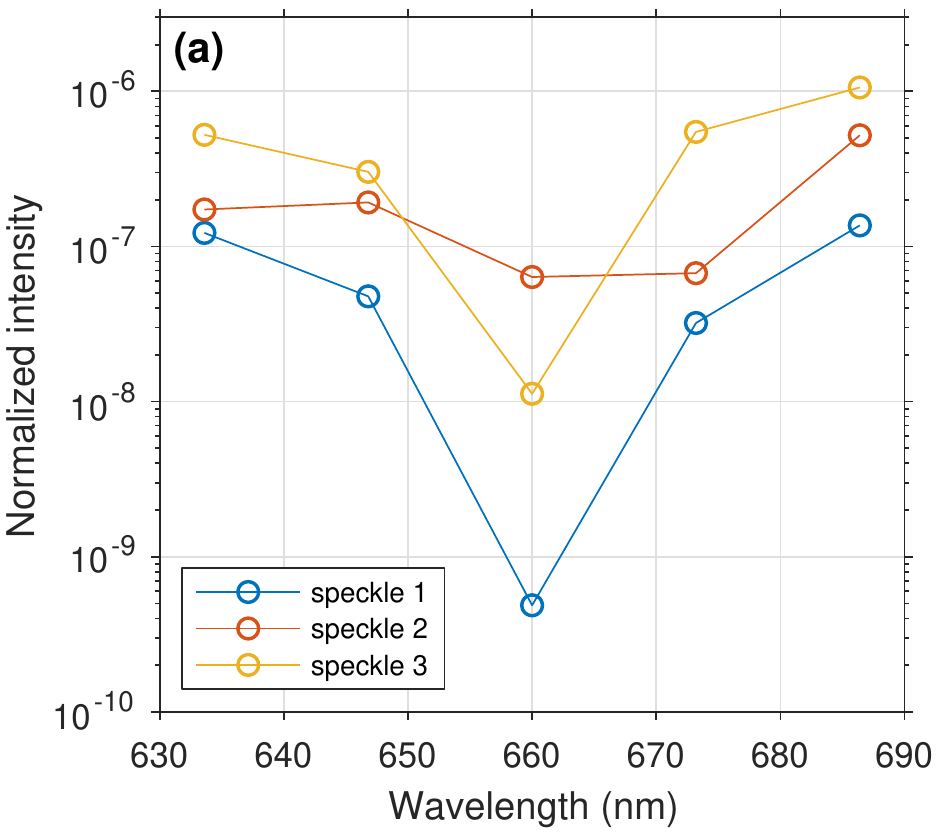}
    \includegraphics[width=0.45\linewidth]{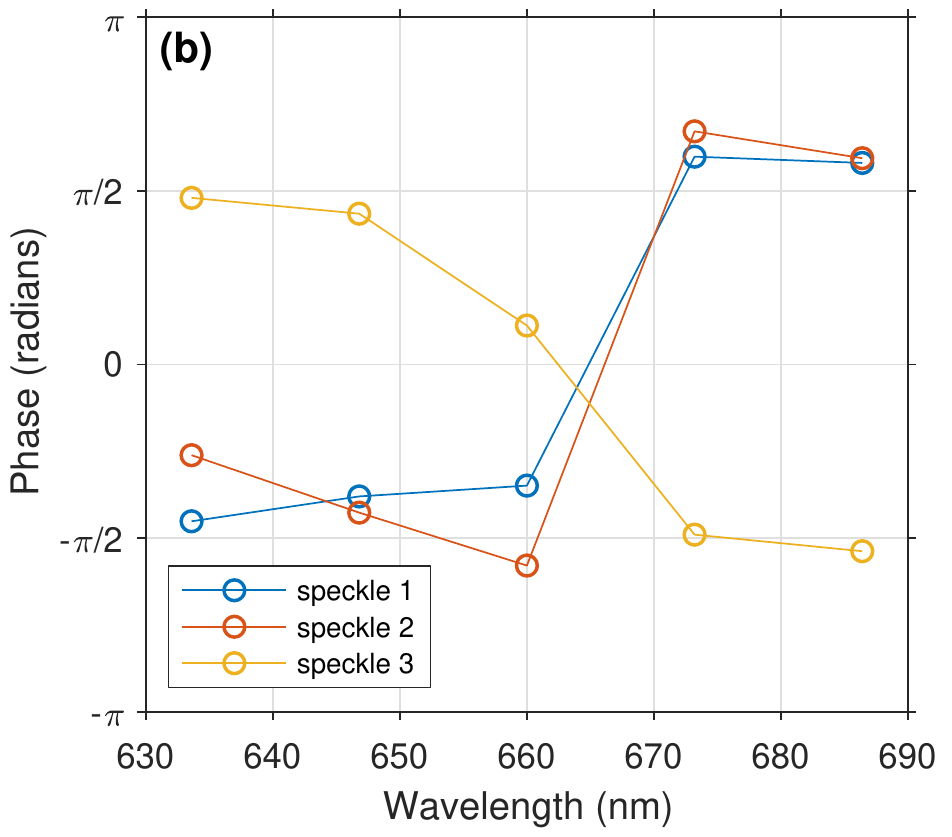}
    \caption{The measured electric field for speckles `1,' `2,' and `3' in Fig.~\ref{fig:Efield_vs_wvl}. \textbf{(a)} Log normalized intensity and \textbf{(b)} phase as at each sub-band.}
    \label{fig:speckles}
\end{figure}

One of the most dominant features in the broadband dark holes were bright localized speckles whose intensity varies strongly with wavelength. We investigated an additional case designed to emphasize these chromatic speckles. Figure~\ref{fig:Efield_vs_wvl} shows the sensed electric field in the image plane from PWP for a 10\% bandwidth, broken up into 5 sub-bands, centered at $\lambda_0$=~660~nm and with a larger dark hole than the previous cases (inner and outer radii of 2~$\lambda_0/D$ and 12~$\lambda_0/D$, respectively). The chromatic speckles were suppressed near the central wavelength, but the wavefront control didn't correct them for all sub-bands simultaneously. After the PWP+EFC converges, these features had a characteristic $\pi$ phase flip indicating a node in the field occurs along the wavelength dimension (see Fig.~\ref{fig:speckles}). We attribute the brightest features to defects in the mask. However, chromatic features are present over the entire dark hole, which may be diffracted from the brightest spots, or due to smaller scale defects in the mask, as seen in Fig.~\ref{fig:retardance_grid}.

\clearpage 

\begin{figure}[t]
    \centering
    \includegraphics[width=0.47\linewidth]{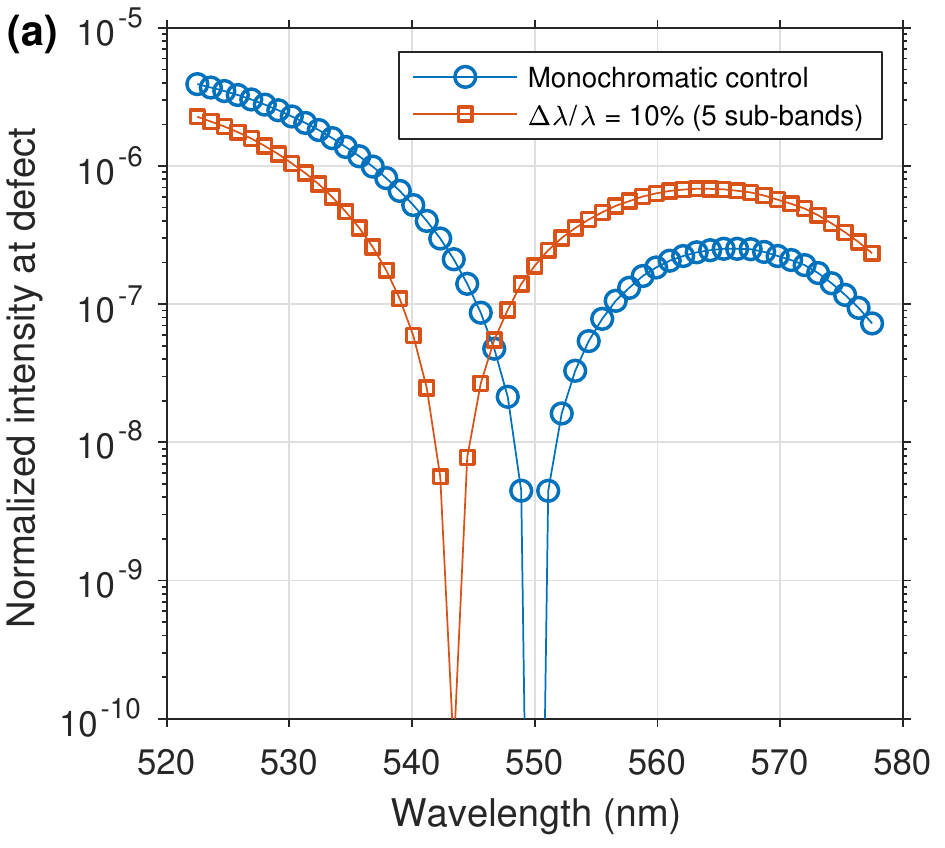}
    \includegraphics[width=0.47\linewidth]{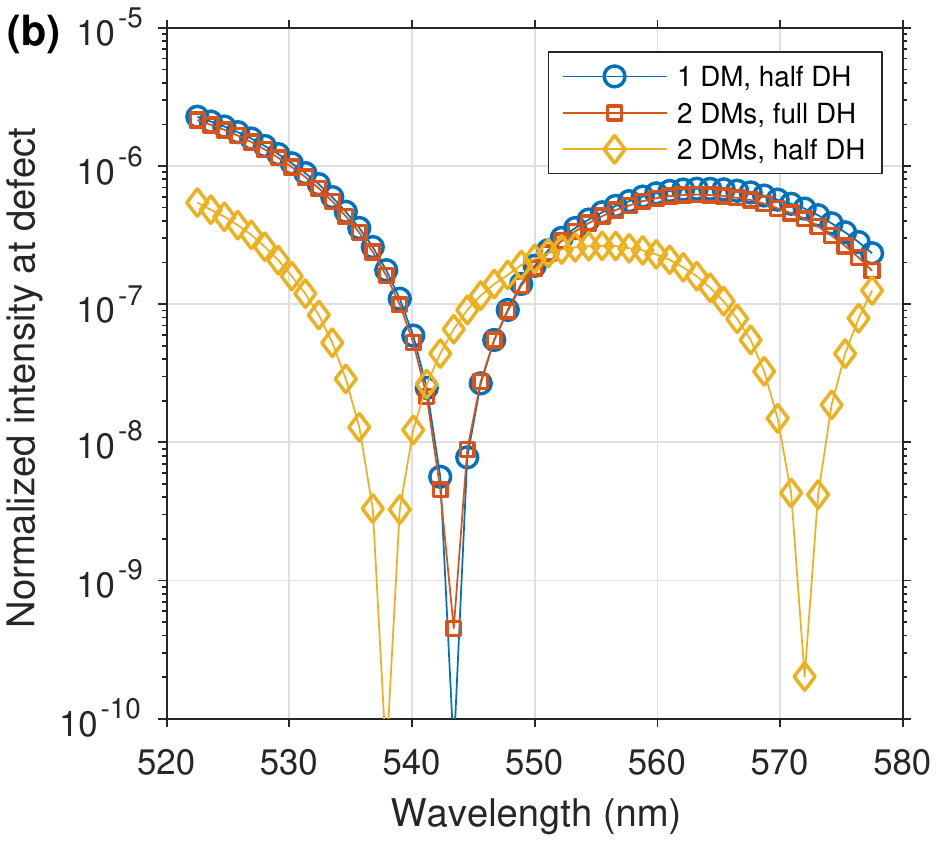}
    \caption{Normalized intensity at the position of a simulated 10~$\mu$m FPM defect separated by 10~$\lambda_0/D$ from the pseudo-star. The wavelength range is a 10\% bandwidth centered at $\lambda_0$~=~550~nm. \textbf{(a)}~Comparison between the EFC result with monochromatic and broadband PWP+EPC control using a single DM. The monochromatic control minimized the intensity at $\lambda_0$~=~550~nm. The broadband control used 5 wavelengths that sampled the 10\% bandwidth. \textbf{(b)}~Comparison between the broadband EFC results using one and two DMs. In the case of two-DM control, correcting only a half dark hole (DH) on one side of the pseudo-star provided additional starlight suppression with two corrected wavelengths within the band. }
    \label{fig:defectmodels}
\end{figure}

\section{Modeling the localized defects}

Our experimental results show that the broadband contrast limitations are related to chromatic speckles that are not removed by the PWP+EFC control algorithm. In this section, we provide evidence that these speckles are likely due to localized manufacturing errors in the mask. 

All of the vortex masks we tested had localized defects in both the retardance and transmittance. The retardance errors are local deviations from the bulk retardance (ideally 180$^\circ$) and the largest retardance errors tend to occur near the central phase singularity. However, past simulation work has shown that errors close the central defect have a small impact on the contrast and can even be blocked by an opaque spot up to $\sim\lambda/D$ in diameter without significantly degrading the contrast performance\cite{Ruane2018_JATIS}. There were also smaller retardance errors further away from the center of the mask that led to a spatially non-uniform leakage term that ultimately is blocked through polarization filtering and effectively introduced a small semi-random amplitude and phase error on the vortex term in the image plane. The masks also had errors in their measured transmittance that appeared as dark spots at random locations that were likely caused by contamination during the manufacturing process. 

The location of the defects on the mask has a significant impact on the performance. The most concerning are amplitude and/or phase errors within $\sim$20-30~$\lambda/D$ of the center. To demonstrate the impact of such errors, we simulated a case where the vortex mask had an opaque spot 10~$\mu$m in diameter at 10~$\lambda_0/D$ from the center. Our simulation is based on the configuration of upcoming experiments on the DST\cite{Patterson2019,Seo2019}, which has a focal ratio of 32 at the FPM and two DMs with 48 actuators across. The DMs are 1~m apart and have 1~mm inter-actuator pitch. The defect creates a bright spot that appears at $\sim$10$^{-6}$ contrast at the defect location in a 10\% bandwidth centered at 550~nm. Figure~\ref{fig:defectmodels} shows the normalized intensity at the center of the defect, after the PWP+EFC control loop converged, computed at 51 wavelengths within the 10\% band. Figure~\ref{fig:defectmodels}a compares the case of monochromatic correction to the broadband correction, which minimizes the mean intensity in 5 sub-bands, using only one DM. Since the speckle intensity is chromatic, the intensity minimum shifted towards shorter wavelengths in the case of broadband correction, but only one wavelength was well corrected in both cases. As such, the broadband contrast only improved by 36\% at the defect location when using broadband control instead of monochromatic control (from 8$\times$10$^{-7}$ to 6$\times$10$^{-7}$). The inability to recover a broadband dark hole was because there is no DM shape that introduces a speckle that is fixed to a position in the image plane at all wavelengths. A sinusoidal surface pattern, for example, introduces a speckle (i.e. an off-axis Airy pattern) that moves outward with a radial separation that is proportional to the wavelength. The result of the PWP+EFC control loop is that the DM applies a pattern similar to a sinusoid, which cancels the speckle introduced by the defect at only one wavelength, but must shift in position for neighboring wavelengths. As the DM-injected speckle moves outward with wavelength, it becomes misaligned with the defect location and the contrast degrades. 

One of the advantages of using the DST for upcoming experiments, instead of the legacy vortex testbed, is that DST was two DMs in series, which provides additional control capability. Figure~\ref{fig:defectmodels}b compares the case of using one DM to two DMs for broadband control. Introducing the second DM had a very small impact when the dark hole was two-sided (i.e. a ``full'' dark hole) because the additional degrees of freedom are used to correct the other side of the image plane. However, using two DMs to create a half dark hole allowed the control algorithm to correct a second wavelength within the specified wavelength range. The result was that the broadband contrast was reduced to 1.5$\times$10$^{-7}$, which represents a factor of 4$\times$ improvement over the single DM case. 

Using similar simulations, we set requirements on the size and density of defects on the vortex masks. Assuming the DST focal ratio, to achieve broadband dark holes with 10$^{-10}$ contrast, the maximum allowable opaque defect size within the central 1~mm region of the mask is $\sim$1~$\mu$m in diameter and any defects should be more than 2~$\lambda/D$ apart, which corresponds to 30~$\mu$m. 

\section{Conclusions and future outlook}

Through our experimental investigation of the broadband performance of vector vortex coronagraphs, we demonstrated that localized defects in the liquid crystal vortex masks are likely to be the limiting factor. The bulk retardance errors were sufficiently suppressed through polarization filtering and were apparently not a dominant contribution to the residual intensity in the dark hole after wavefront control. The focal plane mask defects were particularly problematic because they introduced a bright spot in the image that was fixed to the defect location at all wavelengths. A DM near the entrance pupil plane didn't provide sufficient degrees of freedom to compensate for such an error. Rather, the wavefront control algorithm introduced a sinusoidal pattern on the DM that injected a negative speckle at the defect location for one wavelength. Since a DM-injected speckle moves radially outward with wavelength due to diffraction, the field wasn't cancelled at all wavelengths simultaneously. The result is a highly chromatic residual intensity at the defect location. 

Our team has since procured a new generation of masks with tighter tolerances on the occurrence of defects and overall cleanliness. The new masks will be tested on the state-of-the-art DST in HCIT. The DST is more stable than the legacy vortex testbed and provides two, larger-format (48$\times$48) DMs that may help to further mitigate the residuals due to focal plane mask defects. Our next round of experiments will aim to meet a broadband contrast of better than 10$^{-9}$ in a 10\% bandwidth in the visible regime, paving the way for efficient exoplanet imaging with vector vortex coronagraphs on future space telescopes, such as the HabEx and LUVOIR mission concepts. 

\acknowledgments  
The research was carried out at the Jet Propulsion Laboratory, California Institute of Technology, under a contract with the National Aeronautics and Space Administration (80NM0018D0004). 


\small
\bibliography{Library}   
\bibliographystyle{spiebib}   

\end{document}